\documentclass[11pt]{article}
\usepackage[a4paper,top=80pt,lmargin=80pt,rmargin=80pt,bottom=80pt]{geometry}

\usepackage{graphicx}

\usepackage[]{cite}

\usepackage{amsmath}
\usepackage[]{amssymb}
\usepackage[]{mathrsfs}
\usepackage[]{eucal}
\usepackage[]{tensor}
\usepackage[]{amsthm}
\usepackage[]{bm}

\usepackage{booktabs}
\usepackage{multirow}

\usepackage{colortbl} %%% rowcolor

\usepackage{tikz}
\usepackage{tikz-cd}
\newcommand{\btkz}{\begin{tikzpicture}}
\newcommand{\etkz}{\end{tikzpicture}}

\newcommand{\hdg}{\star} %%% Hodge dual
\newcommand{\df}{\mathrm{d}}
\newcommand{\cl}[1]{\overline{#1}}

\newcommand{\Tr}{\mathrm{Tr}}

\newcommand{\qqd}{\, , \quad}
\newcommand{\bc}{\begin{center}}
\newcommand{\ec}{\end{center}}
\newcommand{\be}{\begin{equation}}
\newcommand{\ee}{\end{equation}}

\newcommand{\RR}{\mathscr{R}}
\newcommand{\FF}{\mathcal{F}}
\newcommand{\GG}{\mathcal{G}}

\newcommand{\ii}{\iota}

\newcommand{\defeq}{\mathrel{\mathop:}=}
\newcommand{\eqdef}{\mathrel{=\mkern-5mu\mathop:}}

\newcommand{\norm}[1]{\left\lVert #1 \right\rVert}

\usepackage{dsfont}
\newcommand{\nn}{\mathds{N}}

\newcommand{\rr}{\mathds{R}}
\newcommand{\cc}{\mathds{C}}

\usepackage[unicode]{hyperref}
\usepackage{xcolor}
\definecolor{pastgreen}{HTML}{669900}
\definecolor{pastblue}{HTML}{336699}
\definecolor{pastred}{HTML}{990000}
\definecolor{linkcol}{HTML}{663333}
\definecolor{graya}{rgb}{0.95,0.95,0.95}
\hypersetup{colorlinks,linkcolor={pastblue},citecolor={pastblue},urlcolor={pastblue}}

\theoremstyle{plain} \newtheorem{tm}{Theorem}[section]
\theoremstyle{plain} \newtheorem{lm}[tm]{Lemma}
\theoremstyle{plain} \newtheorem{defn}[tm]{Definition}
\newcommand{\btm}{\begin{tm}}
\newcommand{\etm}{\end{tm}}
\newcommand{\blm}{\begin{lm}}
\newcommand{\elm}{\end{lm}}
\newcommand{\bdefn}{\begin{defn}}
\newcommand{\edefn}{\end{defn}}

\newcommand{\arrs}[1]{\renewcommand\arraystretch{#1}}

\begin{document}

\begin{flushright}
\texttt{ZTF-EP-26-03}
\end{flushright}

\vspace{20pt}

\bc
{\huge Partial Orderings of Curvature Invariants}

\vspace{25pt}

{\Large Ivica Smoli\'c}

\bigskip

e-mail: ismolic@phy.hr

\bigskip

Department of Physics, Faculty of Science, University of Zagreb

Bijeni\v cka cesta 32, 10000 Zagreb, Croatia

\ec

\vspace{15pt}

\begin{abstract}
We establish a new set of pointwise inequalities that order curvature invariants across various Petrov and Segre types of spacetimes. In arbitrary spacetime dimension, we systematically analyze inequalities among contractions of the Ricci tensor. We further explore the conditions under which all Zakhary--McIntosh invariants in $(1+3)$-dimensional spacetimes are bounded above (up to appropriate powers) by the Kretschmann scalar. These results establish a practical hierarchy among curvature scalars and clarify the extent to which higher-order invariants are algebraically controlled by lower-order ones or vice versa.
\end{abstract} 

\vspace{5pt}


%%%%%%%%%%%%%%%%%%%%%%%%%%
\section{Introduction} %%%
%%%%%%%%%%%%%%%%%%%%%%%%%%

Curvature invariants are scalars constructed from the metric, the Levi--Civita tensor, the Riemann tensor and its covariant derivatives \cite{MacCallum15}. Their history in geometry and topology stretches from Gauss' \emph{Theorema Egregium}, through Riemann's \emph{Habilitationsvortrag}, to modern use in the context of characteristic classes on fibre bundles. The classification of polynomial curvature invariants without covariant derivatives has been studied in gravitational physics for decades, starting with an early paper by Narlikar and Karmarkar \cite{NK49}, aimed at reducing them to a convenient set that always contains a subset of algebraically independent invariants. The smallest known such a set consists of 17 so-called Zakhary--McIntosh (ZM) invariants \cite{ZMc97}. To the best of our knowledge, a complete and general classification of higher-derivative curvature invariants has not yet been established (see e.g.~\cite{McA86}). While numerous syzygies (algebraic relations) are known\footnote{There are several systematic studies of the syzygies among invariants, such as \cite{Harvey95}, and two series of papers, \cite{Sneddon96,Sneddon97,Sneddon99} and \cite{LC04,LC06,LC07}.}, a systematic understanding of inequalities among invariants is still lacking. 

\smallskip

Spacetime singularities are notoriously elusive \cite{Geroch68,Wald,Earman}, with multiple conceptually inequivalent definitions \cite{ES77,ES79}. Motivated by the celebrated singularity theorems \cite{Penrose65,HEll,SG15}, spacetime is denoted as singular if it contains incomplete inextendible geodesics or curves of bounded acceleration (for a recent approach via ``volume incompleteness'' see \cite{GH24}). Unbounded curvature invariants are frequently employed as diagnostics of inextendibility, but this usage is often quite narrow, being limited to the calculation of just a small number of ZM invariants. A thorough analysis can be highly involved, prompting the question of whether some general results could render the task more tractable. Given a spacetime $(M,g_{ab})$ and a set of curvature invariants $\{I_1,\dots,I_n\}$, two fundamental questions, among numerous others, arise:
\begin{itemize}
\item[(1)] Which curvature invariants may be pairwise compared?

\item[(2)] Is there a ``minimal'' invariant which bounds all other invariants?
\end{itemize}

\noindent
Here we set aside nonlocal inequalities, such as those formulated in terms of $L^p$ norms, and focus on pointwise comparisons. Without the qualifier ``minimal'', the second question would have an obvious, straightforward answer with $\mathcal{I} = \sum_i I_i^2$, as $I_i^2 \le \mathcal{I}$ for all $i$. The problem becomes far more challenging and interesting once we look into systematic ordering of the invariants and try to construct simpler upper bound, preferably related to physical observables. Recent work in \cite{DMS25} marked the first steps of this endeavour, establishing inequalities between basic invariants across different families of spacetimes. Some of these inequalities were subsequently generalized in \cite{SKY26}. 

\smallskip

There are two approaches to the given problem: a convenient choice of the coordinate system \cite{DMS25}, or an algebraic decomposition of tensors \cite{SKY26}. Since the latter provides a more efficient and systematic strategy, we shall refine previous results based on the Segre classification  and extend the analysis to the Petrov types of the Weyl tensor. Main tools for the proofs are variations of the H\"older's and classical mean inequalities \cite{HLP,AF,Steele}.

\smallskip

The paper is organized as follows. In Section 2 we briefly summarize definitions of the basic curvature invariants. In Section 3 we present a systematic study of inequalities between contractions of Ricci tensor, clarify the role of the energy conditions and look more closely into the solutions of the Einstein field equations with the ideal fluid or the electromagnetic field. In Section 4 we discuss general properties of ZM invariants from the perspective of Petrov classification, while in Section 5 we prove a set of inequalities between the Kretschmann scalar and ZM invariants for the spherically symmetric, Petrov type D spacetimes. Finally, in Section 6 we summarize the results and discuss remaining open questions. In Appendix A we list definitions of components of the Ricci spinor and the Weyl spinor, together with their properties with respect to spin-boost transformations, and in Appendix B we give definitions of all 17 ZM invariants. Throughout the paper we always assume that the spacetime $(M,g_{ab})$ consists of a smooth, orientable manifold $M$, and a smooth Lorentzian metric $g_{ab}$.

%%%%%%%%%%%%%%%%%%%%%%%%%%%%%%%%%%%%%%%%%%%%%
\section{Elementary curvature invariants} %%%
%%%%%%%%%%%%%%%%%%%%%%%%%%%%%%%%%%%%%%%%%%%%%

We first establish a dictionary of curvature invariants in a general $D$-dimensional spacetime. Unless stated otherwise, we assume $D \ge 2$. The most frequently used are Ricci scalar $R \defeq g^{ab} R_{ab}$ and Kretschmann scalar $K \defeq R_{abcd} R^{abcd}$. The sequence of contracted Ricci tensors is denoted by
\be
\RR_n \defeq \tensor{R}{^{a_1}_{a_2}} \tensor{R}{^{a_2}_{a_3}} \cdots \tensor{R}{^{a_n}_{a_1}}
\ee
for any $n \ge 2$. In particular, we define $\RR_1 \defeq \tensor{R}{^a_a} = R$. The trace-free Ricci tensor is defined by
\be
S_{ab} \defeq R_{ab} - \frac{R}{D} \, g_{ab} ,
\ee
and the associated contractions are
\be
\mathscr{S}_n \defeq \tensor{S}{^{a_1}_{a_2}} \tensor{S}{^{a_2}_{a_3}} \cdots \tensor{S}{^{a_n}_{a_1}}
\ee
for any $n \ge 2$. In particular, we define $\mathscr{S}_1 \defeq \tensor{S}{^a_a} = 0$. Binomial theorem immediately implies
\be
\mathscr{S}_n = \frac{(-R)^n}{D^{n-1}} + \sum_{k=0}^{n-1} \binom{n}{k} (-R/D)^k \, \RR_{n-k} .
\ee
Weyl tensor may be defined for $D \ge 3$ with
\be
C_{abcd} \defeq R_{abcd} - \frac{2}{D-2} \, \big( g_{a[c} R_{d]b} - g_{b[c} R_{d]a} \big) + \frac{2R}{(D-1)(D-2)} \, g_{a[c} g_{d]b} ,
\ee
while its ``square'' is $\mathscr{W} \defeq C_{abcd} C^{abcd}$. Four basic scalars are related by an elegant equation
\be\label{eq:aux}
K = \mathscr{W} + \frac{4}{D-2} \, \RR_2 - \frac{2R^2}{(D-1)(D-2)} .
\ee
In general, none of the invariants, $\RR_n$, $\mathscr{S}_n$, $K$ or $\mathscr{W}$ is either positive or negative definite (see \cite{DMS25,SKY26,Schmidt03} for a detailed discussion).

%%%%%%%%%%%%%%%%%%%%%%%%%%%%%%%%%%%%%%%%%%%%%%%%%%%
\section{Inequalities between Ricci invariants} %%%
%%%%%%%%%%%%%%%%%%%%%%%%%%%%%%%%%%%%%%%%%%%%%%%%%%%

Some of the inequalities between Ricci invariants $\RR_n$ were originally proven \cite{DMS25}, then generalized with Segre classification in \cite{SKY26}. The main point is that
\be\label{eq:ineqR}
\RR_s^{2m} \le D^{2m-s} \RR_{2m}^s
\ee
for all integers $1 \le s < 2m$, given that Ricci tensor belongs to an algebraic type with all real eigenvalues (in $D=4$ case this excludes Segre type A2, that is $[11,Z\cl{Z}]$ or  $[(11),Z\cl{Z}]$, according to the nomenclature in \cite{SKMHH}). Inequalities (\ref{eq:ineqR}) are informing us that, for example, each invariant from the set $\{\RR_1, \dots, \RR_{2m-1}\}$ is bounded from above (with appropriate powers) by $\RR_{2m}$. Still, one might ask if any of these invariants are, in fact, of the ``same order'' in a sense that we have an opposite inequality of the similar form. In order to answer this question we shall prove a theorem with an auxiliary lemma.

\smallskip

\blm
For any finite nonempty set of real numbers $\{x_1,\dots,x_D\}$ and $m,n,p,s \in \nn$, such that $1 \le 2n < p$ and $1 \le s < 2m$, the following inequalities hold
\be
\left( \sum_{i=1}^D x_i^p \right)^{\!2n} \le \left( \sum_{i=1}^D x_i^{2n} \right)^{\!p} , \quad \left( \sum_{i=1}^D x_i^s \right)^{\!2m} \le D^{2m-s} \left( \sum_{i=1}^D x_i^{2m} \right)^{\!s} .
\ee
\elm

\noindent
\emph{Proof}. At the heart of the argument is the equivalence of norms of finite-dimensional normed vector spaces $(\rr^D,\norm{\,.\,}_p)$ and $(\rr^D,\norm{\,.\,}_s)$, encapsulated by the inequalities 
\be
\norm{\bm{x}}_p \le \norm{\bm{x}}_s \le D^{\frac{p-s}{ps}} \norm{\bm{x}}_p ,
\ee
that is
\be\label{ineq:norms}
\left( \sum_{i=1}^D |x_i|^p \right)^{\!s} \le \left( \sum_{i=1}^D |x_i|^s \right)^{\!p} \le D^{p-s} \left( \sum_{i=1}^D |x_i|^p \right)^{\!s} ,
\ee
valid for $1 \le s < p$. If $\bm{x} = \bm{0}$ we have trivial equalities, so let us assume that $\bm{x} \ne \bm{0}$. The first inequality may be proven with introduction of the auxiliary vector $\bm{y} = \bm{x}/\norm{\bm{x}}_s$, such that $\norm{\bm{y}}_s = 1$. Then from $\sum_i |y_i|^s = 1$ it follows that $|y_i| \le 1$ and, since $s < p$, $\sum_i |y_i|^p \le \sum_i |y_i|^s = 1$. Hence, $\norm{\bm{y}}_p \le 1$, which implies $\norm{\bm{x}}_p = \norm{\bm{x}}_s \norm{\bm{y}}_p \le \norm{\bm{x}}_s$. The second inequality follows from the H\"older's inequality (see remarks in \cite{DMS25}). Finally, using (\ref{ineq:norms}) for even $s = 2n$ we get
\be
\left( \sum_{i=1}^D x_i^p \right)^{\!2n} \le \left( \sum_{i=1}^D |x_i|^p \right)^{\!2n} \le \left( \sum_{i=1}^D |x_i|^{2n} \right)^{\!p} = \left( \sum_{i=1}^D x_i^{2n} \right)^{\!p} ,
\ee
while for even $p = 2m$ we get
\be
\left( \sum_{i=1}^D x_i^s \right)^{\!2m} \le \left( \sum_{i=1}^D |x_i|^s \right)^{\!2m} \le D^{2m-s} \left( \sum_{i=1}^D |x_i|^{2m} \right)^{\!s} = D^{2m-s} \left( \sum_{i=1}^D x_i^{2m} \right)^{\!s} ,
\ee
which proves the claim. \qed

\smallskip

\btm\label{tm:ineqTr}
Let $A : \cc^D \to \cc^D$ be a linear operator with real eigenvalues. Then for all $m,n,p,s \in \nn$, such that $1 \le 2n < p$ and $1 \le s < 2m$, the following inequalities hold
\be
(\Tr\,A^p)^{2n} \le (\Tr\,A^{2n})^p , \quad (\Tr\,A^s)^{2m} \le D^{2m-s} (\Tr\,A^{2m})^s .
\ee
\etm

\noindent
\emph{Proof}. Given that $\{\lambda_1,\dots,\lambda_D\}$ are eigenvalues of $A$, standard results in linear algebra \cite{Axler} imply that $\Tr\,A^k = \sum_{i=1}^D \lambda_i^k$ for any $k \in \nn$. By assumption, all eigenvalues are real, and the inequalities follow immediately from the previous lemma. \qed

\medskip

The Ricci tensor with one index raised defines, at each point $p \in M$, a real-linear operator $\tensor{R}{^a_b} : T_p M \to T_p M$. Formally, to apply trace inequalities, we consider the complexified tangent space $T_p M \otimes_\rr \cc$ and the unique complex-linear extension of $\tensor{R}{^a_b}$. Thus, as a corollary of the Theorem \ref{tm:ineqTr} we have the following result.

\btm\label{tm:ineqRR}
Let $(M,g_{ab})$ be a smooth $D$-dimensional Lorentzian manifold. At each point $p \in M$ where all eigenvalues of the Ricci tensor $\tensor{R}{^a_b} : T_p M \to T_p M$ are real, inequalities
\be
\RR_p^{2n} \le \RR_{2n}^p \quad \textrm{and} \quad \RR_s^{2m} \le D^{2m-s} \RR_{2m}^s
\ee
hold for all $m,n,p,s \in \nn$, such that $1 \le 2n < p$ and $1 \le s < 2m$.
\etm 

\noindent
Note that for $2n < p = 2m$ we get $\RR_{2m}^{2n} \le \RR_{2n}^{2m}$, while for $s = 2n < 2m$ we get $\RR_{2n}^{2m} \le D^{2(m-n)} \RR_{2m}^{2n}$. In other words, even Ricci contractions are ``comparable pairs'' in the sense that for each $1 \le 2n < 2m$ the following inequalities hold
\be
\RR_{2m}^{2n} \le \RR_{2n}^{2m} \le D^{2(m-n)} \RR_{2m}^{2n} .
\ee
On the other hand, odd Ricci contractions cannot in general be pairwise compared, but all of them are bounded from above by the even Ricci contractions. Finally, if we rewrite the identity (\ref{eq:aux}) as
\be
(D-1)K = 2\RR_2 + (D-1)\mathscr{W} + \frac{2(D\RR_2 - R^2)}{D-2} ,
\ee
the $(s,m) = (1,1)$ inequality $D\RR_2 \ge R^2 \ge 0$ implies $K \ge \mathscr{W}$. Also, given that $\mathscr{W} \ge 0$ (which is a highly nontrivial assumption, see \cite{Schmidt03}) the $(s,m) = (1,1)$ inequality implies $(D-1)K \ge 2\RR_2$. These inequalities were recently used in proof of a no-go theorem for regular, electrically charged black holes with nonlinear electromagnetic fields \cite{BJS26}. 

\smallskip

All eigenvalues of the trace-free Ricci tensor $\tensor{S}{^a_b}$, with one index raised, are real if and only if the same holds for the Ricci tensor $\tensor{R}{^a_b}$. Thus, under the same assumptions, all inequalities from the Theorem \ref{tm:ineqRR} hold with $\RR_n$ replaced by $\mathscr{S}_n$. Likewise, if the spacetime metric is a solution of the Einstein field equation, $G_{ab} + \Lambda g_{ab} = 8\pi T_{ab}$, then all eigenvalues of the energy-momentum tensor $\tensor{T}{^a_b}$, with one index raised, are real if and only if the same holds for the Ricci tensor $\tensor{R}{^a_b}$. Again, under the same assumptions, all inequalities from the Theorem \ref{tm:ineqRR} hold with $\RR_n$ replaced by contractions $\mathscr{T}_n = \tensor{T}{^{a_1}_{a_2}} \tensor{T}{^{a_2}_{a_3}} \cdots \tensor{T}{^{a_n}_{a_1}}$ of the energy-momentum tensor.

\smallskip

Let us look more closely into the $D=4$ case. The Ricci tensor $\tensor{R}{^a_b}$ has only real eigenvalues if it belongs to Segre types A1, A3 or B (according to the nomenclature in \cite{SKMHH}). The remaining Segre type, A2, may be excluded on physical grounds, specifically by the energy conditions, just as in the case of the Hawking--Ellis type IV energy-momentum tensor \cite{HEll,MMV18}. Suppose for a moment that the energy-momentum tensor $\tensor{T}{^a_b}$ has a complex eigenvalue $\lambda = \alpha + i\beta$, with $\beta \ne 0$, and the corresponding eigenvector $z^a = u^a + iv^a$, such that $\tensor{T}{^a_b} z^b = \lambda z^b$. Then, as the energy-momentum tensor is real and symmetric, we have $(\cl{z}_a \tensor{T}{^a_b} z^b)^* = \cl{z}_a \tensor{T}{^a_b} z^b$ and $(\cl{z}_a \tensor{T}{^a_b} \cl{z}^b)^* = z_a \tensor{T}{^a_b} z^b$. These imply $\beta(u^2 + v^2) = 0$ and $\beta u_a v^a = 0$. Under the assumption $\beta \ne 0$, it follows that $u^2 + v^2 = 0$ and $u_a v^a = 0$. If both $u^a$ and $v^a$ are null, then either $v^a = \kappa u^a$ or $u^a = \kappa v^a$ with $\kappa \in \rr$; in both cases $\tensor{T}{^a_b} z^b = \lambda z^b$ is reduced to $(1+\kappa^2) \beta = 0$, in contradiction with the assumption $\beta \ne 0$. If at least one of the vectors is non-null, the other one must be of the opposite causal type, due to $v^2 = -u^2$. We may, without loss of generality, assume that the timelike vector is future-directed, since the transformation $(u^a,v^a) \mapsto (-u^a,-v^a)$ can always be applied. Here we have two auxiliary future-directed null vectors, either $\ell^a_\pm = u^a \pm v^a$ if $u^a$ is timelike, or $\ell^a_\pm = v^a \pm u^a$ if $v^a$ is timelike. In any case these satisfy $\ell^a_\pm T_{ab} \ell^b_\pm = \pm 2\beta u^2$, which demonstrates that for one of the signs we have a violation of the null energy condition (NEC). Immediately, this implies that the dominant, weak and strong energy conditions are violated. Hence, if the spacetime metric is a solution of the Einstein field equation and the energy-momentum tensor satisfies any of the standard four energy energy conditions (dominant, weak, null or strong), all eigenvalues of the Ricci tensor $\tensor{R}{^a_b}$ are real and previously derived inequalities hold. The basic sample of inequalities for the 4-dimensonal case are listed below,
\bc
\arrs{1.2}
\begin{tabular}{rl}
$(s,m) = (1,1)$ & $\RR_1^2 \le 4 \RR_2$ \\
$(s,m) = (1,2)$ & $\RR_1^4 \le 4^3 \RR_4$ \\
$(s,m) = (2,2)$ & $\RR_2^4 \le 4^2 \RR_4^2$ \\
$(p,n) = (3,1)$ & $\RR_3^2 \le \RR_2^3$ \\
$(s,m) = (3,2)$ & $\RR_3^4 \le 4 \RR_4^3$ \\
$(p,n) = (4,1)$ & $\RR_4^2 \le \RR_2^4$
\end{tabular}
\arrs{1.0}
\ec

\noindent
In order to understand better main points and gain some intuition it is instructive to look at the solutions of the Einstein field equations with the ideal fluid and the electromagnetic field.

%%%%%%%%%%%%%%%%%%%%%%%%%%%%
\subsection{Ideal fluid} %%%
%%%%%%%%%%%%%%%%%%%%%%%%%%%%

The energy momentum tensor of the ideal fluid with density $\rho$, pressure $p$ and fluid element 4-velocity $u^a$ is $T_{ab} = (\rho + p) u_a u_b + p g_{ab}$. Thus, we may conveniently write the Einstein field equation as
\be
\tensor{R}{^a_b} = 4\pi \Big( 2(\rho + p) u^a u_b + (\rho - p) \tensor{g}{^a_b} \Big) .
\ee
In an orthonormal frame $(u^a,e_{(1)}^a,e_{(2)}^a,e_{(3)}^a)$, comoving with the fluid, the Ricci tensor is diagonal, $\tensor{R}{^\alpha_\beta} = 4\pi\,\mathrm{diag}(-\rho-3p,\rho-p,\rho-p,\rho-p)$. Thus, all eigenvalues are real and Ricci tensor belongs to Segre type A1. Also, we immediately have
\be
\RR_n = (4\pi)^n \big( (-1)^n(\rho + 3p)^n + 3(\rho - p)^n \big) .
\ee
Alternatively, for any tensor of the form $\tensor{X}{^a_b} = f u^a u_b + F \tensor{g}{^a_b}$ we have
\be
\tensor{X}{^{a_1}_{a_2}} \tensor{X}{^{a_2}_{a_3}} \cdots \tensor{X}{^{a_n}_{a_1}} = (-f + F)^n + 3 F^n ,
\ee
from where we get contractions of the Ricci tensor. If, in addition, we assume equation of state $p = w\rho$ with some constant $w$, then
\be
\RR_n = (4\pi\rho)^n \big( (-1)^n(1+3w)^n + 3(1-w)^n \big) .
\ee
Contractions of Ricci tensor may be explicitly compared as follows,
\begin{align}
4\RR_2 - \RR_1^2 & = 12(4\pi\rho)^2 (1+w)^2 \\
64 \RR_4 - \RR_1^4 & = 48 (4\pi\rho)^4 (1+w)^2 (5 - 6w + 85w^2) \\
16 \RR_4^2 - \RR_2^4 & = 6144 (4\pi\rho)^8 w^2 (1+w)^2 \big( 1 + 3(4 + 4w + 5w^2)w^2 \big) \\
\RR_2^3 - \RR_3^2 & = 12 (4\pi\rho)^6 (w-1)^2 \big( (5 + 16 w + 20 w^2) + (34 + 48 w + 69 w^2)w^2 \big) \\
4\RR_4^3 - \RR_3^4 & = 48 (4\pi\rho)^{12} (1+w)^2 P(w) \\
\RR_2^4 - \RR_4^2 & = 48 (4\pi\rho)^8 (w-1)^2 (1 + 2w + 5w^2) \big( 5 + 3(14 + 8w + 19w^2)w^2 \big)
\end{align}
with the auxiliary polynomial
\begin{align}
P(w) & = (5 + 2 w + 91 w^2) + (38 + 792 w + 3505 w^2) w^2 + (329 + 3852 w + 11276 w^2) w^4 + \nonumber\\
 & \hspace{15pt} + (12142 + 50328 w + 52 152 w^2) w^6 + (31305 + 63810 w + 32517 w^2) w^8 .
\end{align}
All polynomials are suggestively organized so that it is easy to check that they are nonnegative for any real $\rho$ and $w$. The inequalities are saturated when either $\rho = 0$ or, depending on the specific inequality, if $w = 1$ (stiff matter), $w = 0$ (pressureless dust) or $w = -1$ (cosmological constant). Odd contractions $\RR_1$ and $\RR_3$ in general cannot be compared, as can be seen from the following two examples,
\begin{itemize}
\item[(a)] for $w = 1/3$ (radiation) we have $\RR_1 = 0$ and $\RR_3 = -64(4\pi\rho)^3/9$;
\item[(b)] for $w = (\sqrt[3]{3} - 1)/(\sqrt[3]{3} + 3) \approx 0.1$ we have $\RR_3 = 0$ and $\RR_1 = 16\pi (3 - \sqrt[3]{3})/(3 + \sqrt[3]{3}) \rho$.
\end{itemize}
We note in passing that contractions of the trace-free Ricci tensor have somewhat simpler form,
\be
\mathscr{S}_n = (2\pi)^n (\rho + p)^n (3 + (-3)^n) ,
\ee
allowing a neat identity
\be
\big( 3 + 3^{2n} \big)^m \mathscr{S}_{2m}^n = \big( 3 + 3^{2m} \big)^n \mathscr{S}_{2n}^m
\ee
for all $m,n \in \nn$. Furthermore, the metric and Kretschmann scalar of conformally flat spacetimes filled with an ideal fluid were analysed in \cite{GJ13}.

%%%%%%%%%%%%%%%%%%%%%%%%%%%%%%%%%%%%%%
\subsection{Electromagnetic field} %%%
%%%%%%%%%%%%%%%%%%%%%%%%%%%%%%%%%%%%%%

The energy-momentum tensor of the electromagnetic field $F_{ab}$ may be conveniently written with help of the Hodge dual ${\hdg F}_{ab} \defeq \tensor{\epsilon}{_a_b^c^d} F_{cd}/2$ as
\be
T_{ab} = \frac{1}{8\pi} \, \big( F_{ac} \tensor{F}{_b^c} + {\hdg F}_{ac} \, \tensor{{\hdg F}}{_b^c} \big) .
\ee
As the electromagnetic energy-momentum tensor is traceless, $g^{ab} T_{ab} = 0$, the Einstein field equation is reduced to
\be
\tensor{R}{^a_b} = F^{ac} F_{bc} + {\hdg F}^{ac} \, {\hdg F}_{bc} .
\ee
A revealing insight into the structure of the Ricci tensor comes with help of the identities
\begin{align}
F^{ac} F_{bc} - {\hdg F}^{ac} {\hdg F}_{bc} & = \frac{1}{2} \, \FF \, \tensor{\delta}{^a_c} , \\
F^{ac} {\hdg F}_{bc} = {\hdg F}^{ac} F_{bc} & = \frac{1}{4} \, \GG \, \tensor{\delta}{^a_c} ,
\end{align}
written with the electromagnetic invariants $\FF \defeq F_{ab} F^{ab}$ and $\GG \defeq F_{ab}\,{\hdg F}^{ab}$. Straightforward evaluation leads to
\be
\tensor{R}{^a_b} \tensor{R}{^b_c} = \frac{1}{4} \, (\FF^2 + \GG^2) \, \tensor{\delta}{^a_c} .
\ee
Hence, all Ricci's eigenvalues are real and equal to
\be
\lambda_\pm = \pm \frac{1}{2} \sqrt{\FF^2 + \GG^2} \, .
\ee
If the electromagnetic field is null, $\FF = 0 = \GG$, then all eigenvalues are zero and Ricci tensor belongs to the Segre type A3 (more precisely, $[(11,2)]$). In this case we have trivially $\RR_n = 0$ for all $n \in \nn$. For a non-null electromagnetic field, the multiplicities $m_\pm$ of the eigenvalues $\lambda_\pm$ are constrained by $m_+ + m_- = 4$ (dimensionality) and $m_+ \lambda_+ + m_- \lambda_- = 0$ (tracelessness). These conditions uniquely fix $m_+ = m_- = 2$. Hence, for the non-null electromagnetic field Ricci tensor belongs to the Segre type A1 (more precisely, $[(11)(1,1)]$). Furthermore, as $\lambda_- = -\lambda_+$, all odd contractions vanish, $\RR_{2n+1} = 0$, while even contractions $\RR_{2n} = 4\lambda_+^{2n}$ may be compactly written as
\be
\RR_{2n} = \frac{1}{2^{2(n-1)}} \, (\FF^2 + \GG^2)^n .
\ee
Alternatively, expressions for contractions may be elegantly derived via spinor formalism (see e.g.~\cite{BSJ22}). To summarize, even contractions are pairwise related by
\be
4^m \RR_{2m}^n = 4^n \RR_{2n}^m
\ee
for all $m,n \in \nn$.

%%%%%%%%%%%%%%%%%%%%%%%%%%%%%%%%%%%%%%%%%%%%%%%%%%%%%%%%%%%%%%
\section{Spinors, ZM invariants and Petrov classification} %%%
%%%%%%%%%%%%%%%%%%%%%%%%%%%%%%%%%%%%%%%%%%%%%%%%%%%%%%%%%%%%%%

We focus next on the Zakhary--McIntosh (ZM) invariants in 4-dimensional spacetimes, exploring them within the spinor formalism in which they were first introduced \cite{ZMc97}. Basic elements are the Ricci scalar $R$, the Ricci spinor $\Phi_{ABA'B'}$ and the Weyl spinor $\Psi_{ABCD}$. Traditionally, 17 ZM invariants are divided into four \emph{Weyl invariants} $\{ I_1,I_2,I_3,I_4 \}$, contractions of Weyl spinor, four \emph{Ricci invariants} $\{ I_5,I_6,I_7,I_8 \}$, consisting of $I_5 = R$ and three contractions of Ricci spinor, and nine \emph{mixed invariants} $\{ I_9,\dots,I_{17} \}$, contractions of both Ricci and Weyl spinors. All definitions are gathered in Table \ref{tab:ZM} in Appendix B. One must be wary of the various, and often inconsistent, tensorial reinterpretations of ZM invariants found in the literature, which may differ by normalization or even involve distinct linear combinations.

\smallskip

Spinor space $S$ is a 2-dimensional complex vector space with a nondegenerate 2-form $\epsilon_{AB}$. Its dual space $S^*$ is equipped with the antisymmetric tensor $\epsilon^{AB}$, defined such that $\epsilon^{AB} \epsilon_{BC} = -\tensor{\delta}{^A_C}$. Isomorphisms between elements $\alpha^A$ of the spinor space and elements $\alpha_A$ of its dual space are defined according to $\alpha_A = \alpha^B \epsilon_{BA}$ and $\alpha^A = \epsilon^{AB} \alpha_B$ (``\textbf{l}eft to \textbf{l}ower, \textbf{r}ight to \textbf{r}ise'' index manipulation mnemonic). The whole construction is repeated with their complex conjugate spaces $\cl{S}$ and $\cl{S}^*$ (for details see \cite{PR1,Wald}). By convention, their elements are denoted with bar and primed indices, respectively $\cl{\alpha}^{A'}$ and $\cl{\alpha}_{A'}$ (for the tensors $\epsilon_{A'B'}$ and $\epsilon^{A'B'}$ we omit the bars). A basis of spinor space $S$ is usually denoted by $(o^A,\ii^A)$, normalized such that $\epsilon_{AB} o^A \ii^B = 1$, while the basis of its complex conjugate space $\cl{S}$ is usually denoted by $(\cl{o}^{A'},\cl{\ii}^{A'})$, normalized such that $\epsilon_{A'B'} \cl{o}^{A'} \cl{\ii}^{B'} = 1$. The normalization conditions can also be written as $o_A \ii^A = 1 = -\ii_A o^A$ and $\cl{o}_{A'} \cl{\ii}^{A'} = 1 = -\cl{\ii}_{A'} \cl{o}^{A'}$. For any $\lambda \in \cc^\times$ we may define a new basis element $\tilde{o}^A = \lambda o^A$. To preserve the normalization condition, we complete the basis by defining the other element as $\tilde{\ii}^A = \lambda^{-1} \ii^A$. Thus, we have so-called spin-boost transformation
\be\label{eq:sb}
(o^A,\ii^A) \mapsto (\tilde{o}^A,\tilde{\ii}^A) = (ae^{i\theta} o^A, a^{-1} e^{-i\theta} \ii^A)
\ee
with $a > 0$ and $\theta \in \rr$, which maps old basis $(o^A,\ii^A)$ to a new basis $(\tilde{o}^A,\tilde{\ii}^A)$.

\smallskip

Using conventional components of the Ricci spinor (\ref{eq:compPhi}) and Weyl spinor (\ref{eq:compPsi}), we may decompose them as follows, 
\begin{align}
\Phi_{ABA'B'} & = \Phi_{00} \, \ii_A \ii_B \cl{\ii}_{A'} \cl{\ii}_{B'} - 2\Phi_{01} \, \ii_A \ii_B \cl{\ii}_{(A'} \cl{o}_{B')} + \Phi_{02} \, \ii_A \ii_B \cl{o}_{A'} \cl{o}_{B'} - \nonumber \\
 & \hspace{-12pt} -2\Phi_{10} \, \ii_{(A} o_{B)} \cl{\ii}_{A'} \cl{\ii}_{B'} + 4\Phi_{11} \, \ii_{(A} o_{B)} \cl{\ii}_{(A'} \cl{o}_{B')} - 2\Phi_{12}  \ii_{(A} o_{B)} \cl{o}_{A'} \cl{o}_{B'} + \\
 & + \Phi_{20} \, o_A o_B \cl{\ii}_{A'} \cl{\ii}_{B'} - 2\Phi_{21} o_A o_B \cl{\ii}_{(A'} \cl{o}_{B')} + \Phi_{22} o_A o_B \cl{o}_{A'} \cl{o}_{B'} , \nonumber \\[0.3em]
\Psi_{ABCD} & = \Psi_0 \, \ii_A \ii_B \ii_C \ii_D - 4\Psi_1 \, \ii_{(A} \ii_B \ii_C o_{D)} + 6\Psi_2 \, \ii_{(A} \ii_B o_C o_{D)} - \nonumber \\
 & \hspace{35pt} - 4\Psi_3 \, \ii_{(A} o_B o_C o_{D)} + \Psi_4 \, o_A o_B o_C o_D .
\end{align}
These spinors are related to trace-free Ricci tensor and Weyl tensor via \cite{PR1}
\begin{align}
S_{ab} & = S_{ABA'B'} = -2\Phi_{ABA'B'} , \\
C_{abcd} & = C_{ABCDA'B'C'D'} = \Psi_{ABCD} \epsilon_{A'B'} \epsilon_{C'D'} + \epsilon_{AB} \epsilon_{CD} \cl{\Psi}_{A'B'C'D'} .
\end{align}
It is instructive to notice that
\be
\mathscr{W} = 4 \left( \Psi_{ABCD} \Psi^{ABCD} + \cl{\Psi}_{A'B'C'D'} \cl{\Psi}^{A'B'C'D'} \right) = 48 I_1 .
\ee
Weyl spinor is totally symmetric, $\Psi_{ABCD} = \Psi_{(ABCD)}$, thus by the standard result (see e.g.~Proposition 3.5.18 in \cite{PR1}) there are spinors $(\alpha^A,\beta^A,\gamma^A,\delta^A)$, so-called principal spinors of $\Psi_{ABCD}$, such that 
\be
\Psi_{ABCD} = \alpha_{(A} \beta_B \gamma_C \delta_{D)} .
\ee
Each principal spinor determines a real null vector, $\alpha^{A} \cl{\alpha}^{A'}$ etc., defining a so-called principal null direction of the Weyl tensor. The Petrov classification then follows from the multiplicities of these spinors, as summarized in Table \ref{tab:Pet}. Petrov type I is usually referred to as \emph{algebraically general}, while all others are \emph{algebraically special}.

\bc
\begin{table}[ht]
\centering

\arrs{1.3}
\begin{tabular}{rlll}
Type & $\Psi_{ABCD}$ & $\Psi_i$ & \\
\toprule
I & $\alpha_{(A}\beta_B\gamma_C\delta_{D)}$ & $\Psi_0,\Psi_2,\Psi_4$ & $\mathbb{I}^3 \ne 6\mathbb{J}^2$ \\
\midrule
II & $\alpha_{(A}\alpha_B\gamma_C\delta_{D)}$ & $\Psi_2,\Psi_4$ & \multirow{2}{*}{$\mathbb{I}^3 = 6\mathbb{J}^2 \ne 0$} \\
D & $\alpha_{(A}\alpha_B\beta_C\beta_{D)}$ & $\Psi_2$ & \\
\midrule
III & $\alpha_{(A}\alpha_B\alpha_C\beta_{D)}$ & $\Psi_3$ & \multirow{3}{*}{$\mathbb{I} = \mathbb{J} = 0$} \\
N & $\alpha_A\alpha_B\alpha_C\alpha_D$ & $\Psi_4$ & \\
O & $0$ & $0$ & \\
\end{tabular}
\arrs{1.0}

\caption{Petrov classification. The second column gives the canonical form of the Weyl spinor for each Petrov type, the third column gives the conventional pattern of Weyl spinor components in the chosen spinor basis \cite{PR2}, while fourth column gives properties of the corresponding invariants $\mathbb{I}$ and $\mathbb{J}$.}\label{tab:Pet}
\end{table}
\ec

Weyl spinor may be interpreted as a linear map $\phi_{AB} \mapsto \tensor{\Psi}{_A_B^C^D} \phi_{CD}$ on the 3-dimensional complex space of symmetric spinors. Again, using standard linear algebra result \cite{Axler}, $\mathrm{Tr} A^n = \sum_i \lambda_i^n$ for an operator $A$ with the eigenvalues $\lambda_i$, Weyl spinor eigenvalues $(\lambda_1,\lambda_2,\lambda_3)$ satisfy
\begin{align}
\lambda_1 + \lambda_2 + \lambda_3 & = \tensor{\Psi}{_A_B^A^B} = \Psi_{ABCD} \epsilon^{AC} \epsilon^{BD} = 0 , \\
\lambda_1^2 + \lambda_2^2 + \lambda_3^2 & = \tensor{\Psi}{_A_B^C^D} \tensor{\Psi}{_C_D^A^B} \eqdef \mathbb{I} , \\
\lambda_1^3 + \lambda_2^3 + \lambda_3^3 & = \tensor{\Psi}{_A_B^C^D} \tensor{\Psi}{_C_D^E^F} \tensor{\Psi}{_E_F^A^B} \eqdef \mathbb{J} .
\end{align}
Newly introduced invariants are simple combinations of the first four ZM curvature invariants, $\mathbb{I} = 6(I_1 + iI_2)$ and $\mathbb{J} = 6(I_3 + iI_4)$. More explicitly, expanded in Weyl spinor components $\Psi_i$, they are
\be
\mathbb{I} = 2 (\Psi_0 \Psi_4 - 4\Psi_1\Psi_3 + 3\Psi_2^2) \qqd \mathbb{J} = 6 \begin{vmatrix} \Psi_0 & \Psi_1 & \Psi_2 \\ \Psi_1 & \Psi_2 & \Psi_3 \\ \Psi_2 & \Psi_3 & \Psi_4 \end{vmatrix} .
\ee
Characteristic polynomial of the Weyl spinor is\footnote{The characteristic polynomial $p_A$ of an operator $A : \cc^3 \to \cc^3$ with eigenvalues $(\lambda_1,\lambda_2,\lambda_3)$ can be written as $p_A(\lambda) = \prod_i \, (\lambda - \lambda_i) = \lambda^3 - e_1 \lambda^2 + e_2 \lambda - e_3$, where $e_k = e_k(\lambda_1,\lambda_2,\lambda_3)$ are the elementary symmetric polynomials. By Newton's identities, $e_1 = p_1$, $e_2 = (p_1^2 - p_2)/2$ and $e_3 = (p_1^3 - 3p_1 p_2 + 2p_3)/6$, the coefficients are expressed in terms of the power sums $p_k = \sum_i \lambda_i^k = \mathrm{Tr}\,A^k$.}
\be
p(\lambda) = \lambda^3 - \frac{\mathbb{I}}{2}\,\lambda - \frac{\mathbb{J}}{3} .
\ee
Discriminant of the cubic equation $p(\lambda) = 0$ is
\be
\Delta = (\mathbb{I}^3 - 6\mathbb{J}^2)/2 = (\lambda_1 - \lambda_2)^2 (\lambda_2 - \lambda_3)^2 (\lambda_3 - \lambda_1)^2 .
\ee
Algebraically special Weyl tensors are characterized by a degeneracy of the eigenvalues, equivalently by $\Delta = 0$. This condition is precisely one of the syzygies, $\mathbb{I}^3 = 6\mathbb{J}^2$, which, translated into ZM invariants, reads
\be
(I_1^2 + I_2^2)^3 = (I_3^2 + I_4^2)^2 .
\ee 
In particular, for Petrov types III, N and O we have trivially $\mathbb{I} = \mathbb{J} = 0$, that is $I_1 = I_2 = I_3 = I_4 = 0$. A straightforward choice for an upper bound of the Weyl invariants in Petrov types II and D is to introduce $Z \defeq I_1 + iI_2$, which satisfies $|I_1| \le |Z|$, $|I_2| \le |Z|$, $|I_3|^2 \le |Z|^3$ and $|I_4|^2 \le |Z|^3$.

\smallskip

If the Weyl tensor is Petrov type O, that is $\Psi_{ABCD} = 0$, then the only nontrivial ZM invariants are Ricci invariants, which were already analysed in the previous section (see also remarks about FLRW metric in \cite{DMS25}). The auxiliary relation (\ref{eq:aux}) is reduced to
\be\label{eq:KO}
3K = 2\RR_2 + (4\RR_2 - R^2) .
\ee
Hence, given that the Weyl tensor is Petrov type O \emph{and} Ricci tensor is \emph{not} Segre type A2, by the previously proven inequalities, $4\RR_2 \ge R^2 \ge 0$, we have $3K \ge 2\RR_2$, which implies that Kretschmann scalar bounds all ZM invariants. 

\smallskip

Although the Weyl spinor $\Psi_{ABCD}$ is not identically zero for the Petrov types N and III, all Weyl invariants vanish and the auxiliary relation (\ref{eq:aux}) is again reduced to (\ref{eq:KO}). Thus, given that the Weyl tensor is either Petrov type N or III \emph{and} Ricci tensor is \emph{not} Segre type A2, Kretschmann scalar bounds all Ricci invariants. The challenging aspect of the analysis lies in the mixed invariants, which contain nontrivial components $\Psi_i$ of the Weyl spinors and thus cannot, in general, be bounded by Ricci invariants or the Kretschmann scalar. A notable simplification is that invariants $I_{11}$, $I_{12}$, $I_{16}$ and $I_{17}$ have 2-index contractions of the Weyl spinors, immediately implying that $I_{11}= I_{12} = I_{16} = I_{17} = 0$. Pairs of invariants $(I_9,I_{10})$ and $(I_{13},I_{14})$ are real and imaginary parts of a complex scalar, making direct pairwise comparison generally impossible. Hence, Petrov type N and III mixed invariants can be bounded from above by the sum $I_9^2 + I_{10}^2 + I_{13}^2 + I_{14}^2 + I_{15}^2$. Analysis of the mixed invariants for the Petrov types D, II and I becomes increasingly complex. We shall demonstrate in the following section some of the elegant conclusions that may be achieved under the additional symmetry assumptions.

\smallskip

If the metric is a solution of the vacuum Einstein field equation then $\Phi_{ABA'B'} = 0$ and the only nontrivial ZM invariants are Weyl invariants, which may be bounded from above by $I_1^2 + I_2^2 + I_3^2 + I_4^2$ for any Petrov type (with further reductions in the algebraically special cases), while Kretschmann scalar is reduced to $K = \mathscr{W} = 48 I_1$ (in the presence of the cosmological constant $\Lambda$, we have an additional constant invariant $R = 4\Lambda$ and Kretschmann is shifted to $K = \mathscr{W} - R^2/3$). All ZM invariants for the vacuum Petrov type N solutions, such as pp-waves, are identically zero.

%%%%%%%%%%%%%%%%%%%%%%%%%%%%%%%%%%%%%%%%%%%%%%%%%
\section{Spherically symmetric Petrov type D} %%%
%%%%%%%%%%%%%%%%%%%%%%%%%%%%%%%%%%%%%%%%%%%%%%%%%

In order to gain further insight into the properties of the ZM invariants in Petrov type D spacetimes, we impose additional symmetry assumptions. The class of spherically symmetric spacetimes is among the most comprehensively analysed in the literature and therefore provides a natural starting point for our study. We say that the spacetime is spherically symmetric if its isometry group contains a subgroup isomorphic to the group $SO(3)$, and the orbits of this subgroup are spacelike 2-spheres. At each point of a regular orbit, the isotropy subgroup is isomorphic to $SO(2)$. Accordingly, the tangent space admits a natural $2+2$ decomposition into a spacelike 2-plane tangent to the orbit and a Lorentzian 2-plane orthogonal to it. The isotropy group acts as rotations on the tangent 2-plane and trivially on its orthogonal complement. The induced action is most conveniently described using a Newman–Penrose (NP) null tetrad adapted to this splitting,
\be
\ell^a = o^A \cl{o}^{A'} \qqd n^a = \ii^A \cl{\ii}^{A'} \qqd m^a = o^A \cl{\ii}^{A'} \qqd \cl{m}^a = \ii^A \cl{o}^{A'} .
\ee
The complex null vectors $m^a$ and $\cl{m}^a$ span the spacelike 2-plane tangent to the group orbit, while $\ell^a$ and $n^a$ span the orthogonal Lorentzian 2-plane. The spin-boost transformations (\ref{eq:sb}) have two special cases, boosts for $\theta = 0$, which correspond to $(\ell^a, n^a, m^a, \cl{m}^a) \mapsto (a^2 \ell^a, a^{-2} n^a, m^a, \cl{m}^a)$, and rotations for $a = 1$, which correspond to $(\ell^a, n^a, m^a, \cl{m}^a) \mapsto (\ell^a, n^a, e^{2i\theta} m^a, e^{-2i\theta} \cl{m}^a)$. From (\ref{eq:compPhi}) and (\ref{eq:compPsi}) we see that the only components of the Ricci and the Weyl tensor invariant under arbitrary rotations (i.e.~for all $\theta$) are 
\begin{equation*}
\Phi_{00}, \ \Phi_{11}, \ \Phi_{22} \ \, \textrm{and} \ \, \Psi_2 \, .
\end{equation*}
By construction we already know that $\Phi_{00}$, $\Phi_{11}$ and $\Phi_{22}$ are real. We claim that in the spherically symmetric case the same holds for $\Psi_2$. Namely, a spherically symmetric metric can be expressed as a warped product (see Appendix B in \cite{HEll})
\be
\df s^2 = h_{\mu\nu}(x) \, \df x^\mu \, \df x^\nu + \psi(x) \gamma_{ij}(y) \, \df y^i \, \df y^j ,
\ee
with time-radial coordinates $x^\mu = (t,r)$, angular coordinates $y^i = (\theta,\varphi)$, the Lorentzian metric $h_{\mu\nu}$, the warping function $\psi(x)$, and the standard metric $\gamma_{ij}$ on the unit 2-sphere. It is a well-known result that the mixed Weyl tensor components may be decomposed according to $C_{\mu i \nu j} = H_{\mu\nu}(x) \gamma_{ij}(y)$. As a consequence, contraction $C_{abcd} \ell^a m^b \cl{m}^c n^d = -H_{\mu\nu} \ell^\mu n^\nu \gamma_{ij} m^i \cl{m}^j$ is real, which implies that $\Psi_2$ is real as well. The resulting ZM invariants are straightforward, but tedious to evaluate\footnote{A simple computational trick for spinor contractions is to write the spinor basis explicitly as vectors. For example, we can define 
\texttt{uo = (1,0)}, \texttt{ui = (0,1)}, \texttt{do = (0,1)}, and \texttt{di = (-1,0)} to represent, respectively, $o^A$, $\ii^A$, $o_A$, and $\ii_A$, which can be implemented in symbolic or numerical languages (e.g., as lists in \emph{Mathematica} or arrays in \emph{Python}), allowing ordinary vector operations to handle spinors without extra formalism.}: some of them, $\{I_2,I_4,I_{10},I_{12},I_{14},I_{17}\}$, are identically zero, while all nontrivial, apart from $I_5 = R$, are listed in the Table \ref{tab:sphPetD}.

\bc
\begin{table}[ht]
\centering

\arrs{1.3}
\begin{tabular}{rlccc}
$i$ & $I_i$ & $\iota_i$ & $1/C_i$ & $\kappa_i$ \\ [3pt]
$1$ & $\Psi_2^2$ & $1$ & $2^4 \cdot 3$ & $1$ \\ \rowcolor{graya}
$3$ & $-\Psi_2^3$ & $2$ & $2^{12} \cdot 3^3$ & $3$ \\
$6$ & $2(\Phi_{00} \Phi_{22} + 2\Phi_{11}^2)/3$ & $1$ & $2^3 \cdot 3$ & $1$ \\ \rowcolor{graya}
$7$ & $2\Phi_{00} \Phi_{11} \Phi_{22}$ & $2$ & $2^9 \cdot 3^3$ & $3$ \\
$8$ & $\Phi_{00} \Phi_{22} (\Phi_{00} \Phi_{22} + 8\Phi_{11}^2)/3$ & $1$ & $2^6 \cdot 3^2$ & $2$ \\ \rowcolor{graya}
$9$ & $2\Psi_2 (\Phi_{00} \Phi_{22} - 4\Phi_{11}^2)$ & $2$ & $2^6 \cdot 3^4$ & $3$ \\
$11$ & $-2\Psi_2^2 (\Phi_{00} \Phi_{22} - 4\Phi_{11}^2)$ & $2$ & $2^{16} \cdot 3^2$ & $4$ \\ \rowcolor{graya}
$13$ & $-2\Psi_2 \Phi_{00} \Phi_{22} (\Phi_{00} \Phi_{22} - 4\Phi_{11}^2)$ \hspace{5pt} & $2$ & $2^{10} \cdot 3 \cdot 5^5$ & $5$ \\
$15$ & $2\Psi_2^2 (\Phi_{00} \Phi_{22} + 8\Phi_{11}^2)$ & $1$ & $2^7 \cdot 3$ & $2$ \\ \rowcolor{graya}
$16$ & $-2\Psi_2^3 (\Phi_{00} \Phi_{22} + 8\Phi_{11}^2)$ & $2$ & $2^{12} \cdot 5^5$ & $5$
\end{tabular}
\arrs{1.0}

\caption{Nontrivial ZM invariants, excluding $I_5 = R$, for the spherically symmetric Petrov type D metric. Last three columns contain, respectively, the invariant powers $\iota_i$, the reciprocal constants $1/C_i$ (expressed as a product of prime factors for clarity) and the Kretschmann powers from the inequality (\ref{ineq:ICK0}).}\label{tab:sphPetD}
\end{table}
\ec

Our next goal is to compare ZM invariants with the Kretschmann scalar, which may be expressed from (\ref{eq:aux}),
\be
K = \mathscr{W} + 2\mathscr{S}_2 + \frac{1}{6} \, R^2 .
\ee
Taking into account that in general $\mathscr{W} = 48 I_1$ and $\mathscr{S}_2 = 12 I_6$, while for spherically symmetric Petrov type D (see Table \ref{tab:sphPetD}) $I_1 = \Psi_2^2$ and $I_6 = 2(\Phi_{00}\Phi_{22} + 2\Phi_{11}^2)/3$, we have
\be
K = 48\Psi_2^2 + 16 (\Phi_{00}\Phi_{22} + 2\Phi_{11}^2) + \frac{1}{6} \, R^2 .
\ee
As the Ricci scalar $R$ appears explicitly only in $I_5$, it is convenient to introduce ``reduced Kretschmann scalar'',
\be
K_0 \defeq K - \frac{1}{6} \, R^2 = 16 (3\Psi_2^2 + \Phi_{00}\Phi_{22} + 2\Phi_{11}^2) .
\ee
The main result of this section is summarized in the following theorem and the accompanying lemma.

\blm
Let $x_i > 0$, $n_i \in \nn_0$ and $a_i > 0$ for $i \in \{1,\dots,m\}$, such that $N = \sum_i n_i \ge 1$. Then
\be\label{ineq:xCax}
\prod_{i=1}^m x_i^{n_i} \le C \left( \sum_{i=1}^m a_i x_i \right)^{\!N} , \ \textrm{with} \ \ C = \prod_{i=1}^m \left( \frac{n_i}{N a_i} \right)^{\!n_i}
\ee
and convention that $0^0 = 1$.
\elm

\noindent
\emph{Proof}. We start from the weighted AM-GM inequality\footnote{An elegant proof \cite{Steele} of this classical result rests upon elementary inequality $x \le e^{x-1}$ valid for all $x \in \rr$ (the equality holds iff $x=1$). Using the abbreviation $s = \sum_k w_k r_k$, we have $r_i/s \le e^{(r_i - s)/s}$ and $(r_i/s)^{w_i} \le e^{w_i(r_i - s)/s}$ for all $i$. Finally, by multiplying all these inequalities, we get $\prod_i (r_i/s)^{w_i} \le e^0 = 1$, which immediately leads to (\ref{ineq:AMGM}).}
\be\label{ineq:AMGM}
\prod_{i=1}^m r_i^{w_i} \le \sum_{i=1}^m w_i r_i ,
\ee
valid for positive real numbers $r_i > 0$ and weights $w_i \ge 0$, normalized such that $\sum_i w_i = 1$. Equality holds iff all $r_i$ with $w_i \ne 0$ are equal. Let us denote by $V \subseteq \{1,\dots,m\}$ the subset of indices for which $n_i \ne 0$. Now we insert $w_i = n_i/N$ in (\ref{ineq:AMGM}) and note that for all $i \notin V$ we have $r_i^{w_i} = 1$ and $w_i r_i = 0$, so that these terms are effectively absent from the inequality. Finally, for all $i \in V$ we use the substitution $r_i = a_i x_i N/n_i$ and raise both sides to power $N$, which leads us to
\be
\prod_{i=1}^m x_i^{n_i} = \prod_{i \in V} x_i^{n_i} \le \left( \prod_{i\in V} \left( \frac{n_i}{N a_i} \right)^{\!n_i} \right) \left( \sum_{i \in V} a_i x_i \right)^{\!N} \le C \left( \sum_{i=1}^m a_i x_i \right)^{\!N} .
\ee
As by convention $0^0 = 1$, we have a simplification $C = \prod_{i\in V} (n_i/N a_i)^{n_i}$. \qed

\smallskip

\btm
Let $(M,g_{ab})$ be spherically symmetric, Petrov type D spacetime, with the Ricci tensor which satisfies null convergence condition, $R_{ab} \ell^a \ell^b \ge 0$ for all null vectors $\ell^a$. Then the nontrivial ZM invariants $I_i$ for $i \ne 5$ are bounded by the reduced Kretschmann scalar $K_0$ with the inequality 
\be\label{ineq:ICK0}
I_i^{\iota_i} \le C_i K_0^{\kappa_i} ,
\ee
where powers $\iota_i, \kappa_i > 0$ and constants $C_i > 0$, are listed in the Table \ref{tab:sphPetD}.
\etm

\noindent
\emph{Proof}. First technical difficulty lies with the term $\Phi_{00}\Phi_{22}$ which, in general, is not necessarily nonnegative. However, as in the language of NP null tetrad
\be
4\Phi_{00}\Phi_{22} = (S_{ab} \ell^a \ell^b)(S_{cd} n^c n^d) ,
\ee
we see that the null convergence condition for the Ricci tensor implies $\Phi_{00}\Phi_{22} \ge 0$. Hence, we can divide analysis into the case A, when $\Phi_{00}\Phi_{22} > 0$, and the case B, when $\Phi_{00}\Phi_{22} = 0$. In the case A we may use the remaining boost transformation,
\begin{equation*}
(\Phi_{00},\Phi_{11},\Phi_{22}) \mapsto (a^2 \Phi_{00}, \Phi_{11}, a^{-2} \Phi_{22})
\end{equation*}
to enforce further simplification $\Phi_{22} = \Phi_{00}$.

\smallskip

Furthermore, practical choice of the curvature invariant powers in (\ref{ineq:ICK0}) is $\iota_i = 1$ when $I_i$ is manifestly nonnegative and $\iota_i = 2$ otherwise. Then, using abbreviations $x = \Psi_2^2$, $y = \Phi_{00}^2$ and $z = \Phi_{11}^2$, we see that the problem is reduced to comparison of the polynomials $(I_i(x,y,z))^{\iota_i}$ with the polynomial $K_0(x,y,z) = 16(3x + y + 2z)$. Each monomial $x^{n_1} y^{n_2} z^{n_3}$ can be compared with $K_0$ using inequality (\ref{ineq:xCax}). Since each polynomial $I_i$ is homogeneous, this comparison consistently determines the corresponding powers $\kappa_i$, which are listed in Table \ref{tab:sphPetD}. In the special case B, invariants $\{I_7,I_8,I_{13}\}$ are identically zero and inequalities collapse to trivial claims.

\smallskip

Finally, we address the question of the optimal constants $C_i$. Initial estimate is obtained by a simple sum of the individual monomial inequalities of the form $A x^{n_1} y^{n_2} z^{n_3} \le C K_0^N$. Sharp constants for each inequality are defined by
\be
C_i = \sup_D \frac{I_i^{\iota_i}}{K_0^{\kappa_i}} ,
\ee
where supremum is over the domain $D = \{(x,y,z) \in \rr^3 \mid x,y,z \ge 0, \, xyz \ne 0 \}$. In other words, $C_i$ is the smallest positive constant for which the inequality $I_i^{\iota_i} \le C_i K_0^{\kappa_i}$ holds on the whole domain $D$. The key advantage in computation of this supremum is that $I_i^{\iota_i}$ and $K_0^{\kappa_i}$ are homogeneous polynomials of the same degree $d_i$, namely
\be
(I_i(\lambda x, \lambda y, \lambda z))^{\iota_i} = \lambda^{d_i} (I_i(x,y,z))^{\iota_i} , \ (K_0(\lambda x, \lambda y, \lambda z))^{\kappa_i} = \lambda^{d_i} (K_0(x,y,z))^{\kappa_i}
\ee
for any $\lambda > 0$. Using this freedom, we may reduce the problem with the choice of the constraint $3x + y +2z = 1$ and then investigate the constrained supremum of the numerator $I_i^{\iota_i}$. In particular, one must check for local maxima in the interior and on the boundary of the constrained domain. Results, presented as reciprocal sharp constants, are listed in Table \ref{tab:ZM}. We note that sharp constants agree between cases A and B for all nontrivial cases, $i \in \{1,3,6,9,11,15,16\}$. \qed

\medskip

Once (\ref{ineq:ICK0}) is established, $0 \le K_0 \le K$ implies that the inequalities $I_i^{\iota_i} \le C_i K^{\kappa_i}$ hold for all $i \ne 5$ and we have, trivially, $I_5^2 \le 6K$. To summarize, all Zakhary--McIntosh invariants are bounded above by the Kretschmann scalar, with appropriate powers, for the spherically symmetric, Petrov type D spacetimes, given that the Ricci tensor satisfies null convergence condition. If the spacetime is a solution of the Einstein field equation, we can give another view on the conclusions from the previous two sections.

\btm
Let $(M,g_{ab})$ be a spacetime with spherically symmetric metric $g_{ab}$ which is a solution of the Einstein field equation $G_{ab} + \Lambda g_{ab} = 8\pi T_{ab}$ and the energy momentum tensor $T_{ab}$ satisfies null energy condition. Then all ZM invariants are bounded by the Kretschmann scalar by inequalities $I_i^{\iota_i} \le C_i K^{\kappa_i}$, for some powers $\iota_i, \kappa_i > 0$ and constants $C_i > 0$.
\etm

Namely, null energy conditions immediately implies null convergence condition and forbids complex eigenvalues of the Ricci tensor $\tensor{R}{^a_b}$. Furthermore, spherically symmetric spacetime is either Petrov type O or D, both of which have been discussed above in detail.

%%%%%%%%%%%%%%%%%%%%%%%%
\section{Discussion} %%%
%%%%%%%%%%%%%%%%%%%%%%%%

Inequalities provide a powerful tool to systematically organize the intricate web of curvature invariants. The success of this programme is prominent with the contractions of the Ricci tensors, treated in Section 3. If the Ricci tensor $\tensor{R}{^a_b}$ has all eigenvalues real, we have two practical implications: (a) if one of contractions $\RR_n$ (either even or odd) is unbounded, then all other even contractions must be unbounded, and (b) if one even contraction $\RR_{2k}$ is bounded, then all other contractions $\RR_n$ are bounded. In $(1+3)$-dimensional case, assuming the metric is a solution of the Einstein field equations, the eigenvalues of $\tensor{T}{^a_b}$, and hence of $\tensor{R}{^a_b}$, are real whenever the energy-momentum tensor satisfies any of the four standard energy conditions (dominant, weak, null, or strong). Finer details of underlying inequalities, in case when spacetime contains either ideal fluid or electromagnetic field, are discussed in subsections 3.1 and 3.2.

\smallskip

More ambitious part of the analysis is focused on the whole set of ZM curvature invariants. In Section 4 we have given a broad overview of the problem, with some partial resolutions, from the perspective of Petrov classification. The simplest case is Petrov type O, in which the only nontrivial ZM invariants are Ricci invariants: under the additional assumption that the Ricci tensor $\tensor{R}{^a_b}$ has all eigenvalues real, inequalities from the Theorem \ref{tm:ineqRR} hold and all Ricci invariants are bounded above by the Kretschmann scalar. Petrov types N and III are still relatively manageable, as the Weyl invariants and some of the mixed invariants are identically zero, but for the Petrov types D, II and I the problem becomes increasingly involved. In Section 5 we have delved deeper into the special case of spherically symmetric Petrov type D, proving a set of inequalities (\ref{ineq:ICK0}). Again, this provides us with two practical implications (under the given assumptions): (a) if any of the ZM invariants is unbounded, then so is the Kretschmann scalar, and (b) if the Kretschmann scalar is bounded, then so are all ZM invariants.

\smallskip

Kretschmann scalar is often informally equated with the tidal forces, although this relation is somewhat subtle. The tidal acceleration $a^a$ for a family of geodesics with timelike 4-velocity $u^a$ and deviation (Jacobi) vector field $X^a$ is governed by the geodesic deviation (Jacobi) equation $a^a = \tensor{R}{^a_b_c_d} u^b u^c X^d$. On the other hand, we can define the electric and the magnetic parts of the Weyl tensor with respect to the given geodesics \cite{McIAWH94,MB98}, $E_{ab} = C_{acbd} u^c u^d$ and $B_{ab} = {\hdg C}_{acbd} u^c u^d$, where the Hodge dual of Weyl tensor is defined as ${\hdg C}_{abcd} = \tensor{\epsilon}{_a_b^p^q} C_{pqcd}/2$. Electric tensor $E_{ab}$ and magnetic tensor $B_{ab}$ are symmetric and spacelike, in sense that $u^a E_{ab} = 0$ and $u^a B_{ab} = 0$. In analogy with the decomposition of the electromagnetic invariants into electric and magnetic contributions, two quadratic Weyl invariants can be written as $48I_1 = \mathscr{W} = 8(E_{ab} E^{ab} - B_{ab} B^{ab})$ and $48 I_2 = C_{abcd}\,{\hdg C}^{abcd} = 16 E_{ab} B^{ab}$. If the spacetime is a vacuum solution of the Einstein field equation, we have a simple relation $a_a = -E_{ab} X^b$ and, since $a^a$ and $E_{ab}$ are spacelike, we can use Cauchy--Bunyakovsky--Schwarz inequality to obtain a bound $a_a a^a \le E_{bc} E^{bc} X_d X^d$. Finally, in case when the Weyl tensor is purely electric (i.e.~$B_{ab} = 0$), we have $\mathscr{W} = K$ and the inequality $8a_a a^a \le K X_b X^b$, which bounds tidal acceleration with Kretschmann scalar and norm of the deviation vector. Of course, this still leaves the possibility of bounded tidal acceleration with unbounded Kretschmann scalar (for a discussion, see \cite{ORGSP16,MRLOC24,RGetal25,ZG25}). When any of the underlying assumptions are not satisfied, for instance due to the presence of matter or a nonvanishing magnetic component of the Weyl tensor, the bound requires careful generalization.

\smallskip

This leaves us with several open questions, the most important being
\begin{itemize}
\item[(1)] What are physically interesting upper bounds for the ZM invariants in Petrov type III and N spacetimes?
\item[(2)] How to generalizes inequalities for the Petrov type D spacetimes beyond spherical symmetry?
\item[(3)] How to systematically study inequalities for the Petrov type I and II spacetimes?
\end{itemize}

\noindent
Their resolution will help us better understand the role of curvature invariants in the characterization of spacetime and, hopefully, elucidate the relationships between disparate definitions of singularities.

%%%%%%%%%%%%%%%%%%%%%%%%%%%%%%%%%%%%%%%%%%%%%%%%%%%%%%%%%%
%%%%%%%%%%%%%%%%%%%%%%%%%%%%%%%%%%%%%%%%%%%%%%%%%%%%%%%%%%
\section*{Acknowledgements}
The author would like to thank Diego Rubiera-Garcia for numerous discussions about spacetime singularities. The research was supported by the European Union --- NextGenerationEU through the National Recovery and Resilience Plan 2021-2026 and the Croatian Science Foundation Project No. IP-2025-02-8625, \emph{Quantum aspects of gravity}. 

%%%%%%%%%%%%%%%%%%%%%%%%%%%%%%%%%%%%%%%%%%%%%%%%%%%%%%%%%%
%%%%%%%%%%%%%%%%%%%%%%%%%%%%%%%%%%%%%%%%%%%%%%%%%%%%%%%%%%

%%%%%%%%%%%%%
%%%%%%%%%%%%%
\appendix %%%
%%%%%%%%%%%%%
%%%%%%%%%%%%%

%%%%%%%%%%%%%%%%%%%%%%%%%%%%%%%%%%%%%%%%%%%%%%%
\section{Ricci and Weyl components} %%%
%%%%%%%%%%%%%%%%%%%%%%%%%%%%%%%%%%%%%%%%%%%%%%%

Definitions of the Ricci spinor components and corresponding transformations with respect to the spin-boost transformation (\ref{eq:sb}),  
\begin{align}\label{eq:compPhi}
\Phi_{00} & \defeq \Phi_{ABA'B'} o^A o^B \cl{o}^{A'} \cl{o}^{B'} & & \widetilde{\Phi}_{00} = a^4 \Phi_{00} \nonumber \\
\Phi_{01} & \defeq \Phi_{ABA'B'} o^A o^B \cl{o}^{A'} \cl{\ii}^{B'} & & \widetilde{\Phi}_{01} = a^2 e^{2i\theta} \Phi_{01} \nonumber \\
\Phi_{02} & \defeq \Phi_{ABA'B'} o^A o^B \cl{\ii}^{A'} \cl{\ii}^{B'} & & \widetilde{\Phi}_{02} =  e^{4i\theta} \Phi_{02} \nonumber \\
\Phi_{10} & \defeq \Phi_{ABA'B'} o^A \ii^B \cl{o}^{A'} \cl{o}^{B'} & & \widetilde{\Phi}_{10} = a^2 e^{-2i\theta} \Phi_{10} \nonumber \\
\Phi_{11} & \defeq \Phi_{ABA'B'} o^A \ii^B \cl{o}^{A'} \cl{\ii}^{B'} & & \widetilde{\Phi}_{11} =  \Phi_{11} \\
\Phi_{12} & \defeq \Phi_{ABA'B'} o^A \ii^B \cl{\ii}^{A'} \cl{\ii}^{B'} & & \widetilde{\Phi}_{12} =  a^{-2} e^{2i\theta} \Phi_{12} \nonumber \\
\Phi_{20} & \defeq \Phi_{ABA'B'} \ii^A \ii^B \cl{o}^{A'} \cl{o}^{B'} & & \widetilde{\Phi}_{20} =  e^{-4i\theta} \Phi_{20} \nonumber \\
\Phi_{21} & \defeq \Phi_{ABA'B'} \ii^A \ii^B \cl{o}^{A'} \cl{\ii}^{B'} \nonumber & & \widetilde{\Phi}_{21} = a^{-2} e^{-2i\theta} \Phi_{21} \\
\Phi_{22} & \defeq \Phi_{ABA'B'} \ii^A \ii^B \cl{\ii}^{A'} \cl{\ii}^{B'} & & \widetilde{\Phi}_{22} = a^{-4} \Phi_{22} \nonumber
\end{align}
Hermiticity of the Ricci spinor implies that $\cl{\Phi}_{ji} = \Phi_{ij}$. Hence, $\Phi_{00}$, $\Phi_{11}$ and $\Phi_{22}$ are real. Definitions of the Weyl spinor components and corresponding transformations with respect to the spin-boost transformation (\ref{eq:sb}), 
\begin{align}\label{eq:compPsi}
\Psi_0 & \defeq \Psi_{ABCD} \, o^A o^B o^C o^D & & \widetilde{\Psi}_0 = a^4 e^{4i\theta} \Psi_0 \nonumber \\
\Psi_1 & \defeq \Psi_{ABCD} \, o^A o^B o^C \ii^D \nonumber & & \widetilde{\Psi}_1 = a^2 e^{2i\theta} \Psi_1 \\
\Psi_2 & \defeq \Psi_{ABCD} \, o^A o^B \ii^C \ii^D & & \widetilde{\Psi}_2 = \Psi_2 \\
\Psi_3 & \defeq \Psi_{ABCD} \, o^A \ii^B \ii^C \ii^D \nonumber & & \widetilde{\Psi}_3 = a^{-2} e^{-2i\theta} \Psi_3 \\
\Psi_4 & \defeq \Psi_{ABCD} \, \ii^A \ii^B \ii^C \ii^D & & \widetilde{\Psi}_4 = a^{-4} e^{-4i\theta} \Psi_4 \nonumber
\end{align}

\newpage

%%%%%%%%%%%%%%%%%%%%%%%%%%%%%%%%%%%%%%%%%%%%%%%
\section{ZM invariants in spinor formalism} %%%
%%%%%%%%%%%%%%%%%%%%%%%%%%%%%%%%%%%%%%%%%%%%%%%

The Zakhary--McIntosh curvature invariants were defined in the original paper \cite{ZMc97} (up to relabeling the indices) as follows:

\bc
\begin{table}[ht]
\centering

\arrs{1.3}
\begin{tabular}{rl}
$I_1$ & $\textrm{Re} \big( \Psi_{ABCD} \Psi^{ABCD} \big)/6$ \\
$I_2$ & $\textrm{Im} \big( \Psi_{ABCD} \Psi^{ABCD} \big)/6$ \\
$I_3$ & $\textrm{Re} \big( \Psi_{ABCD} \tensor{\Psi}{^C^D_E_F} \Psi^{EFAB} \big)/6$ \\
$I_4$ & $\textrm{Im} \big( \Psi_{ABCD} \tensor{\Psi}{^C^D_E_F} \Psi^{EFAB} \big)/6$ \\ [3pt]
$I_5$ & $R$ \\
$I_6$ & $\big( \Phi_{ABA'B'} \Phi^{ABA'B'} \big)/3$ \\
$I_7$ & $\big( \Phi_{ABA'B'} \tensor{\Phi}{^B_C^{B'}_{C'}} \Phi^{ACA'C'} \big)/3$ \\
$I_8$ & $\Phi_{ABC'D'} \tensor{\Phi}{^B_E^{C'}_{F'}} \big( \Phi^{AGH'D'} \tensor{\Phi}{^E_G^{F'}_{H'}} + \Phi^{AGH'F'} \tensor{\Phi}{^E_G^{D'}_{H'}} \big)/6$ \\ [3pt]
$I_9$ & $\textrm{Re} \big( \Psi_{ABCD} \tensor{\Phi}{^C^D_{E'}_{F'}} \Phi^{ABE'F'} \big)$ \\
$I_{10}$ & $\textrm{Im} \big( \Psi_{ABCD} \tensor{\Phi}{^C^D_{E'}_{F'}} \Phi^{ABE'F'} \big)$ \\
$I_{11}$ & $\textrm{Re} \big( \tensor{\Psi}{_(_A_B^E^F} \Psi_{CD)EF} \tensor{\Phi}{^A^B_{G'}_{H'}} \Phi^{CDG'H'} \big)$ \\
$I_{12}$ & $\textrm{Im} \big( \tensor{\Psi}{_(_A_B^E^F} \Psi_{CD)EF} \tensor{\Phi}{^A^B_{G'}_{H'}} \Phi^{CDG'H'} \big)$ \\
$I_{13}$ & $\textrm{Re} \big( \Psi_{ABCD} \tensor{\Phi}{^A_E_{A'}_{B'}} \tensor{\Phi}{^B^E^{A'}_{C'}} \big( \tensor{\Phi}{^C_F^{B'}_{D'}} \Phi^{DFD'C'} + \tensor{\Phi}{^C_F^{C'}_{D'}} \Phi^{DFD'B'} \big) \big)/2$ \\
$I_{14}$ & $\textrm{Im} \big( \Psi_{ABCD} \tensor{\Phi}{^A_E_{A'}_{B'}} \tensor{\Phi}{^B^E^{A'}_{C'}} \big( \tensor{\Phi}{^C_F^{B'}_{D'}} \Phi^{DFD'C'} + \tensor{\Phi}{^C_F^{C'}_{D'}} \Phi^{DFD'B'} \big) \big)/2$ \\
$I_{15}$ & $\tensor{\Psi}{^A^B_E_F} \tensor{\Phi}{^E^F_{C'}_{D'}} \tensor{\cl{\Psi}}{^{C'}^{D'}_{G'}_{H'}} \tensor{\Phi}{_A_B^{G'}^{H'}}$ \\
$I_{16}$ & $\textrm{Re} \big( \tensor{\Psi}{^(^A^B_C_D} \Psi^{EF)CD} \cl{\Psi}_{G'H'I'J'} \tensor{\Phi}{_A_B^{I'}^{J'}} \tensor{\Phi}{_E_F^{G'}^{H'}} \big)$ \\
$I_{17}$ & $\textrm{Im} \big( \tensor{\Psi}{^(^A^B_C_D} \Psi^{EF)CD} \cl{\Psi}_{G'H'I'J'} \tensor{\Phi}{_A_B^{I'}^{J'}} \tensor{\Phi}{_E_F^{G'}^{H'}} \big)$
\end{tabular}
\arrs{1.0}

\caption{ZM invariants.}\label{tab:ZM}
\end{table}
\ec

Invariants $I_{13}$ and $I_{14}$ may be written in a slightly clearer form,
\be
I_{13} + iI_{14} = -\Psi_{ABCD} \Phi^{AEA'B'} \tensor{\Phi}{^B_E_{A'}_{C'}} \Phi^{CFC'D'} \tensor{\Phi}{^D_F_{D'}_{B'}} .
\ee

\newpage

%%%%%%%%%%%%%%%%%%%%%%%%%%%%
%%%%%%%%%%%%%%%%%%%%%%%%%%%%
\bibliographystyle{amsalpha}
\bibliography{poci}
%%%%%%%%%%%%%%%%%%%%%%%%%%%%
%%%%%%%%%%%%%%%%%%%%%%%%%%%%

\end{document}